\documentclass[10pt,journal,compsoc]{IEEEtran}

\usepackage{listings}
\usepackage[utf8]{inputenc}
\usepackage{amsmath}
\usepackage{xcolor}
\usepackage{graphicx}
\usepackage{svg}
\usepackage{url}
\usepackage{caption}
\usepackage{booktabs}
\usepackage{blindtext}
\usepackage{lipsum}

\usepackage{bm}

\definecolor{codegreen}{rgb}{0,0.6,0}
\definecolor{codegray}{rgb}{0.5,0.5,0.5}
\definecolor{codepurple}{rgb}{0.58,0,0.82}
\definecolor{backcolour}{rgb}{1,1,1}

\usepackage{color}
\usepackage{ulem}
\usepackage{comment}
\newcommand{\KTadd}[1]{{\textcolor{black}{#1}}}

\newcommand{\STadd}[1]{{\textcolor{black}{#1}}}

\lstdefinestyle{mystyle}{
    breakatwhitespace=false,    
    breaklines=true,        
    captionpos=b,               
    keepspaces=true,     
    basicstyle=\ttfamily\scriptsize,
    linewidth=8.7cm,
    xleftmargin=0.5cm,
    numbers=left,               
    numbersep=5pt,          
    showspaces=false,           
    showstringspaces=false,
    showtabs=false,         
    tabsize=2,
    frame=lrtb
}
 
\usepackage{enumitem}
\setlist[itemize]{leftmargin=5mm}

\let\origthelstnumber\thelstnumber
\makeatletter
\newcommand*\Suppressnumber{%
  \lst@AddToHook{OnNewLine}{%
    \let\thelstnumber\relax%
     \advance\c@lstnumber-\@ne\relax%
    }%
}

\newcommand*\Reactivatenumber{%
  \lst@AddToHook{OnNewLine}{%
   \let\thelstnumber\origthelstnumber%
   \advance\c@lstnumber\@ne\relax}%
}

\makeatother

\ifCLASSOPTIONcompsoc
  \usepackage[nocompress]{cite}
\else
  \usepackage{cite}
\fi

\begin{document}

\noindent\fbox{%
    \parbox{\textwidth}{%
        © 2021 IEEE.  Personal use of this material is permitted.  Permission from IEEE must be obtained for all other uses, in any current or future media, including reprinting/republishing this material for advertising or promotional purposes, creating new collective works, for resale or redistribution to servers or lists, or reuse of any copyrighted component of this work in other works.
    }%
}

\newpage

\title{PyQUBO: \KTadd{Python Library} for Mapping Combinatorial Optimization Problems\\ to QUBO Form}

\author{Mashiyat~Zaman,
        Kotaro~Tanahashi,
        and~Shu~Tanaka% <-this % stops a space
\IEEEcompsocitemizethanks{\IEEEcompsocthanksitem M. Zaman and K. Tanahashi are with Recruit Communications Co., Ltd., Chuo-ku, Tokyo, 104-0032. E-mail:\{mashiyat\_zaman, tanahashi\}@r.recruit.co.jp\protect\\%
\IEEEcompsocthanksitem S. Tanaka is with Department of Applied Physics and Physico-Informatics, Keio University, Kanagawa 223-8522, Japan, with Green Computing Systems Research Organization, Waseda University, Tokyo, 162-0042 Japan, and with Precursory Research for Embryonic Science and Technology, Japan Science and Technology Agency, Kawaguchi-shi, 332-0012 Japan.
E-mail: shu.tanaka@appi.keio.ac.jp}
}

\date{
     $^1$Recruit Communications Co., Ltd., Chuo, Tokyo 104-0054, Japan \\
     $^2$Department of Applied Physics and Physico-Informatics, Keio University, Kanagawa 223-8522, Japan\\
     $^3$Green Computing Systems Research Organization, Waseda University, Shinjuku, Tokyo 162-0042, Japan \\
     $^4$JST PRESTO, Kawaguchi, Saitama 332-0012, Japan
}

\markboth{}%
{Shell \MakeLowercase{\textit{et al.}}: Bare Demo of IEEEtran.cls for Computer Society Journals}

\IEEEtitleabstractindextext{%
\begin{abstract}
We present PyQUBO, an open-source, Python \KTadd{library} for constructing quadratic unconstrained binary optimizations (QUBOs) from the objective functions and the constraints of optimization problems. PyQUBO enables users to prepare QUBOs or Ising models for various combinatorial optimization problems with ease thanks to the abstraction of expressions and the extensibility of the program. QUBOs and Ising models formulated using PyQUBO are solvable by Ising machines, including quantum annealing machines. We introduce the features of PyQUBO with applications in the number partitioning problem, knapsack problem, graph coloring problem, and integer factorization using a binary multiplier. Moreover, we demonstrate how PyQUBO can be applied to production-scale problems through integration with quantum annealing machines. Through its flexibility and ease of use, PyQUBO has the potential to make quantum annealing a more practical tool among researchers.
\end{abstract}

% Abbreviate Section to Sec., except at the beginning of the section.

}

\maketitle

\IEEEdisplaynontitleabstractindextext

\IEEEpeerreviewmaketitle

\IEEEraisesectionheading{\section{Introduction}\label{sec:introduction}}

\IEEEPARstart{C}{ombinatorial} optimization is the calculation of the maxima or minima of a function within a discrete domain. Various combinatorial optimization problems, such as schedule and shift planning, delivery, and traffic flow, exist in daily life. Such problems often contain numerous possible solutions, making an exhaustive search intractable and hence creating an increasing demand for more efficient technology.

To overcome such computational limitations, a new type of computation technology known as the Ising machine was developed. In 2011, the first commercial quantum annealing machine was presented\cite{johnson2011quantum}. The hardware of existing quantum annealing machines has been developed based on the theories of quantum annealing\cite{kadowaki1998quantum} and adiabatic quantum computation \cite{farhi2000quantum, farhi2001quantum}. Ising machines are inspired not only by quantum annealing but also other principles that have been developed since the emergence of the first commercial quantum annealer \cite{yamaoka201520k,inagaki2016coherent,mcmahon2016fully,aramon2019physics,maezawa2019toward,goto2019combinatorial}. A number of studies utilizing Ising machines have been conducted in various fields: portfolio optimization\cite{rosenberg2016solving}, traffic optimization \cite{neukart2017traffic}, rectangle packing optimization\cite{terada2018ising}, item listing optimization for e-commerce websites\cite{nishimura2019item}, and materials design\cite{kitai2020designing}.

To use Ising machines in solving a problem, the energy function of Ising model or quadratic unconstrained binary optimization (QUBO) corresponding to the objective function and constraints of the problem must be prepared. Here, we refer to the energy function as Hamiltonian. However, programming Ising models and QUBOs for Ising machines may be challenging when the objective function and constraints are complicated. Thus, we developed PyQUBO as \KTadd{a Python library} for programming QUBOs and Ising models. Using PyQUBO's high-level class objects, users can construct Hamiltonians intuitively.
Not only does PyQUBO make it easier to read and write code, but it also makes a program more extensible, thereby allowing users to solve combinatorial optimization problems more efficiently.
Through its accessibility and extensibility, PyQUBO has the potential to make quantum annealing more common among researchers across a wide range of fields.

In this paper, we demonstrate how PyQUBO can be used to express QUBO and the Hamiltonian of Ising model as easily readable Python code. The remainder of this paper is organized as follows. In Section 2, we introduce the Ising machine and explain how we use one to solve a combinatorial optimization problem. In Section 3, we formulate the combinatorial optimization problem, and in Section 4, we demonstrate how it can be expressed in terms of a QUBO or Ising model. In Section 5, we introduce PyQUBO and explore the motivation for its development. In Section 6, we demonstrate how combinatorial optimization problems can be solved by simply using the fundamentals of PyQUBO. In Section 7, we present advanced PyQUBO methods for writing and debugging complex problems. In Section 8, we explain PyQUBO for logical gates. In Section 9, we address the use of PyQUBO in conjunction with the D-Wave Ocean System software and D-Wave Advantage quantum annealing machine. \KTadd{In section 10, we present internal implementations of PyQUBO and benchmark the performance with different implementations including other packages. Section 11 is devoted to the conclusion of the paper.}

\section{Use of Ising Machines}
\label{how_to_use}

In this section, we explain how Ising machines are used to solve combinatorial optimization problems. In general, five steps are involved in using Ising machines to solve optimization problems \cite{tanaka2020theory}, as follows:

\begin{enumerate}
    \item Discern a combinatorial optimization problem from the issue.
    \item Represent the combinatorial optimization problem using an Ising model.
    \item Embed the Ising model into the Ising machine according to the hardware specifications and determine the hyperparameters.
    \item Search for the low-energy states of the Ising model.
    \item Interpret the final state to obtain feasible solutions to the original combinatorial optimization problem.
\end{enumerate}

First, we need to identify the combinatorial optimization problem from the issue in question. Next, we formulate the optimization problem as an Ising model or QUBO. In this process, constraint terms are introduced to the Hamiltonian to satisfy the  constraints. If the original optimization problem contains non-binary discrete variables, such as integer variables, these must be encoded using binary variables. The details of this process are explained in Sections \ref{comb_opt} and \ref{qubo_ising}. 

Third, we map our logical Ising model onto the physical Ising model, which consists of variables and interactions between variables that are implemented on the Ising machine. This mapping process is known as embedding. Because embedding is itself a combinatorial optimization problem, several efficient embedding algorithms have been proposed \cite{cai2014practical,choi2008minor,boothby2016fast}. We also need to specify the hyperparameters, that is, the coefficients of the constraint terms. Fourth, we use the Ising machine to obtain the low-energy states of the physical Ising model. Finally, we interpret the variable states from the Ising machine and obtain the states corresponding to the logical Ising model. We determine whether the solution satisfies the problem constraints: if not, we repeat step 3 and update the hyperparameters\footnote{Depending on the problem structure and how constraints are broken, it is possible to restore to a solution that satisfies constraints \cite{kanamaru2019mapping}.}. Through these five steps, we can obtain feasible solutions to combinatorial optimization problems using the Ising machine. 

\section{Combinatorial Optimization}
\label{comb_opt}

The mathematical formulation for a combinatorial optimization problem is expressed as follows:
\begin{align}
\label{eq:comb_opt}
&\boldsymbol{z}^{*} = \mathop{\mathrm{arg\,min}}\nolimits_{\boldsymbol{z}} f(\boldsymbol{z}),\quad \boldsymbol{z}\in \mathcal{S},\\ \label{eq:comb_opt}
&\begin{cases} g_{\ell}(\boldsymbol{z}) = 0 &\text{($\ell = 1, \ldots, L$)},\\ h_{m}(\boldsymbol{z}) \leq 0 &\text{($m = 1, \ldots, M$)}, \end{cases}\nonumber
\end{align}
where $\boldsymbol{z}$ represents discrete integer decision variables of which number is $n$, ${f(\boldsymbol{z})}$ is the cost function, and $\mathcal{S}$ is the set of decision variables satisfying the given equality and inequality constraints \(g_{\ell}(\boldsymbol{z})\) and \(h_{m}(\boldsymbol{z})\). 
 
Equation (\ref{eq:comb_opt}) can be rewritten as an optimization problem without any constraints using the penalty function method. Given the equality constraint \(g(\boldsymbol{z}) = 0\), we can consider the equation
\begin{equation} 
\label{eq:comb_opt_equality}
\boldsymbol{z}^{*} = \mathop{\mathrm{arg\,min}}\nolimits_{\boldsymbol{z}} \{f(\boldsymbol{z}) + \lambda [g(\boldsymbol{z})]^{2}\},\quad \boldsymbol{z}\in \mathcal{Z}^{n}. 
\end{equation}
Similarly, given the inequality constraint \(h(\boldsymbol{z}) \leq 0\), Eq. (\ref{eq:comb_opt}) can be rewritten as 
\begin{equation} 
\label{eq:comb_opt_inequality}
\boldsymbol{z}^{*} = \mathop{\mathrm{arg\,min}}\nolimits_{\boldsymbol{z}} \{f(\boldsymbol{z}) + \lambda \max[h(\boldsymbol{z}),0]\},\quad \boldsymbol{z}\in \mathcal{Z}^{n}. 
\end{equation}
In both equations, \(\boldsymbol{z}\in \mathcal{Z}^{n}\) indicates that the decision variables must be integers. For sufficiently large values of the coefficient \(\lambda\), Eqs. (\ref{eq:comb_opt_equality}) and (\ref{eq:comb_opt_inequality}) produce feasible solutions that satisfy the constraints with greater probability.
\section{The QUBO and Ising Model}
\label{qubo_ising}

Ising machines use the QUBO or Hamiltonian of the Ising model to solve combinatorial optimization problems. The Ising model and QUBO are defined on an undirected graph \(G=(V, E)\), where $V$ and $E$ are the sets of vertices and edges on $G$, respectively.

\subsection{Ising Model}

The Hamiltonian of the Ising model on $G$ is expressed by 
\begin{equation} 
\label{eq:hamiltonian-graph}
H_{\rm Ising}(\boldsymbol{s}) = \sum_{i \in V} h_{i} s_{i} + \sum_{(ij) \in E} J_{ij} s_{i} s_{j},\quad s_{i} \in \{-1, 1\}, \end{equation}
where \(s_{i}\) is the decision variable called spin at \(i \in V\), \(h_{i}\) is the magnetic field at \(i \in V\), and \(J_{ij}\) is the interaction at the edge $i, j$. Here $h_i$ and $J_{ij}$ are real numbers.

\subsection{QUBO}
\begin{figure}[b]
    \centering
    \includegraphics[width=4cm]{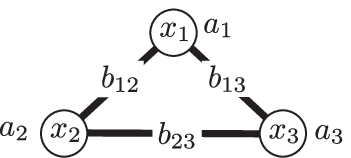}
    \caption{Simple 3-binary variable system indicating weights \(a_{i}\) and interaction strengths \(b_{ij}\). 
    %The weights and strengths can also be represented as the matrix \(Q_{ij}\).
    }
    \label{fig:qubo}
\end{figure}

The QUBO represents the cost function of a binary combinatorial optimization problem with linear and quadratic terms. Let $x_i$ be the $i$-th binary variable. Given the graph $G$, it is formulated as
\begin{align} 
H_{\rm QUBO}(\boldsymbol{x}) = \sum_{i \in V} a_{i} x_{i} + \sum_{(ij) \in E} b_{ij} x_{i} x_{j},\quad x_{i} \in \{0, 1\}
%f(\boldsymbol{x}) &= \sum_{i \in V} a_{i} x_{i} + \sum_{(ij) \in E} b_{ij} x_{i} x_{j} \\
% &= \sum_{i \leq j} Q_{ij} x_{i} x_{j}, 
\end{align}
where $a_i$ and $b_{ij}$ are real numbers. Here $b_{ij}=b_{ji}$ for arbitrary $i$ and $j$. A QUBO defined on the undirected graph \(G=(V,E)\) is illustrated in Fig. \ref{fig:qubo}.
%, where the linear and quadratic coefficients of matrix $Q$ are represented as the weights and interactions, respectively.

Let us confirm the equivalence between QUBO and Ising model. Using the relation \(x_{i} = (s_{i} +1)/2\), QUBO can be used to represent the combinatorial optimization problem in terms of the Ising model. The coefficients of Ising model are given by 
\begin{align}
\label{eq:a_to_h}
h_{i} &= \frac{a_{i}}{2} + \sum_{j \in \partial_{i}} \frac{b_{ij}}{2},\quad \forall i \in V,\\ 
\label{eq:b_to_j}
J_{ij} &= \frac{b_{ij}}{4},\quad \forall (ij) \in E, \end{align}
where $\partial_i$ indicates the set of vertices connected to vertex $i$ by the edges.
By substituting Eqs.~(\ref{eq:a_to_h}) and (\ref{eq:b_to_j}) into Eq.~(\ref{eq:hamiltonian-graph}), we confirm the equivalence between QUBO and Ising model except for a constant value.

QUBO can be expressed in matrix form. Let $Q_{ij}$ be a $|V| \times |V|$ matrix whose elements are given by
\begin{align}
Q_{ij} = 
    \begin{cases}
        a_i & [i=j, \forall i \in V]\\
        b_{ij} & \text{[$\forall(ij) \in E$ and $i < j$]}\\
        0 & \text{[otherwise]}
    \end{cases}.
%Q_{ii} &= a_{i}\quad [\forall i \in V],\\
%Q_{ij} &= \begin{cases} b_{ij} + b_{ji} &\text{[$\forall(ij) \in E$ and $i < j$]}\\ 
%0 &\text{[$\forall(ij) \notin E$, $i \neq j$]}\end{cases}.
\end{align}
By using $Q_{ij}$ and a column vector $\boldsymbol{x}$ generated by arranging binary variables, $H_{\rm QUBO}(\boldsymbol{x})$ is rewritten by
\begin{align}
H_{\rm QUBO}(\boldsymbol{x}) &= \sum_{i \leq j} Q_{ij} x_{i} x_{j} = \boldsymbol{x}^{\rm T} Q \boldsymbol{x},
\end{align}
where $\boldsymbol{x}^{\rm T}$ represents the transpose of vector $\boldsymbol{x}$.
%The QUBO matrix $Q$ and the linear and quadratic coefficients \(a_i\) and \(b_{ij}\) are related by

\subsection{Combinatorial Optimization Problem represented by Ising Model or QUBO}

The equality and inequality constraints described by Eqs. (\ref{eq:comb_opt_equality}) and (\ref{eq:comb_opt_inequality}) must be added as penalty terms that are also written as QUBOs. That is, the Hamiltonian of Ising model or QUBO can be generalized as
\begin{equation} 
\label{eq:generalized}
H = {H}_{\text{cost}} + \lambda {H}_{\text{const}}, \end{equation}
where \(\lambda\) determines the constraint term weight, \({H}_{\text{cost}}\) is the cost function, and \({H}_{\text{const}}\) is the penalty term, which is $0$ when the constraint is satisfied and greater than $0$ otherwise. Methods to construct ${H}_{\text{const}}$ are explained in \cite{lucas2014ising,tanaka2017quantum,tanahashi2019application}.

In summary, a combinatorial optimization problem can be made solvable by a Ising machines by representing the cost function and constraints of the original problem using the linear and quadratic terms of the binary variables. 

\vspace{-3mm}
\section{Introduction to PyQUBO}
\label{intro-to-pyqubo}
Thus far, we have observed that the general combinatorial optimization problem needs to be formulated as QUBO (or an Ising model) to solve the problem using Ising machines. Practically, we take the following steps to obtain the QUBO corresponding to the optimization problem.

\begin{enumerate}
\item Formulate the problem as an integer programming (IP) problem.
\item Reformulate the optimization problem without constraints by introducing constraint terms to the objective function.
\item {\bf Encode the integer variables with binary variables.}
\item {\bf Expand the objective function.}
\item {\bf Reduce the degree of higher-order terms.}
\item {\bf Obtain the QUBO matrix from the coefficient of the polynomial.}
\end{enumerate}

In the first step, we formulate the general combinatorial optimization problem presented in Eq. (\ref{eq:comb_opt}) as an IP problem. The formulations for various combinatorial optimization problems as IP problems have been studied in depth \cite{wolsey1999integer}. This process is also required when we solve the combinatorial optimization problem with general optimization solvers, such as Gurobi \cite{gurobi}.
In the second step, we introduce the constraint terms into the objective function by using the penalty method, so that the optimization problem does not include any additional conditions (see Section \ref{comb_opt}).
In the third step, the integer variable in the objective function are encoded with binary variables. As there are several means of encoding an integer variable and each type has different characteristics, we need to select the appropriate one carefully.
In the fourth step, the objective function is expanded into a sum of products. If the products in the expanded polynomial have a degree greater than two, the order of the products needs to be reduced by introducing new variables and constraint terms into step 5.
Finally, we combine the like terms of the expression to obtain the coefficient of QUBO.

It is necessary to use a program to conduct steps (3)–(6). However, to implement such a program, we need to calculate the expanded form of the objective function in advance, which is usually prepared by hand. Furthermore, when the objective function consists of several complicated terms, the program becomes complicated, which may result in software bugs. When we attempt to solve the problem with Ising machines, we generally try to create several formulation and encoding types for improved results. Thus, if we can transform the problem into QUBO rapidly, we can create different formulations and obtain superior results more efficiently.

To facilitate the creation of QUBOs, we developed a software tool known as PyQUBO. Using PyQUBO, one can define the Hamiltonian, that is, the objective function, with variable objects in a generic format. By calling the compile method of the expression, the QUBO matrix can be obtained instantly, without knowing the expanded form of the Hamiltonian (Section \ref{sec:pyqubo_fundamentals}). If the order of the products in the polynomial exceeds two, the PyQUBO compiler automatically reduces the order and produces the corresponding QUBO matrix. PyQUBO provides several class modules, including multi-dimensional arrays (Supplementary Section A), integer classes with different encodings (Section \ref{subsec:integer}), and logical gates (Section \ref{subsec:logical_gate}). These classes enable not only complicated expressions to be implemented quickly and easily but also the construction of modules on top of these (Section \ref{subsec:binary_multiplier}).
As explained in step 5 of Section \ref{how_to_use}, the solutions need to be validated. PyQUBO also provides a feature to validate the solutions automatically (Section \ref{subsec:constraint}).
\vspace{-3mm}
\subsection{Quick Reference}
% Reorder the questions in order of the sections.
A quick reference to each section, according to what the users wish to accomplish with PyQUBO, is provided below.\\

\STadd{
\vspace{-20pt}
\begin{itemize}
\item {\bf Automatic creation of QUBO}: PyQUBO automatically expands the terms of the Hamiltonian to produce the QUBO matrix, which is compilation (Sec.\ref{model-and-solver}). In addition, PyQUBO automatically reduces the order of the polynomials during compilation (Sec. \ref{sec:order_reduction}).
\item {\bf Validation of constraint satisfaction}: By using the \texttt{Model.decode\_sample()} method, one can verify whether the given solutions are valid (Sec. \ref{decode_solution}).
\item {\bf Update of specific variables}: By defining a value with \texttt{Placeholder} when creating the Hamiltonian, the value we want to change can be specified even after compilation (Sec. \ref{subsec:placeholder}).
\item {\bf Usage of integer decision variables instead of continuous decision variables}: Continuous variables can be approximately represented using the \texttt{Integer} class. For example, the continuous value $x \in [0,1]$ is approximated by $\hat{x} \in \{0,0.1,\dots,1.0\}$. In PyQUBO, $\hat{x}$ can be defined as \texttt{x=0.1*LogEncInteger("x", 0, 10)}. Other \texttt{Integer} class types are also available (Supplementary Section B).
\item {\bf Usage of categorical variables}: In certain problems, such as the graph coloring problem \cite{o2016toq}, a discrete variable is used to represent the category $C \in \{C_1,C_2,\dots,C_n\}$. This variable is known as a categorical variable. In PyQUBO, categorical variables can be defined using the \texttt{OneHotEncInteger} class. Refer to Supplementary Section B for further details.
\item {\bf Usage of integer variables when inequality constraints exist}: PyQUBO provides various types of integer classes that can be used to create inequality constraints (Sec. \ref{knapsack}).
\item {\bf Usage of logical variables}: PyQUBO provides logical gate classes and logical gate constraint classes, which are useful for formulating the satisfiability problem (SAT) or integer factoring as QUBOs (Sec. \ref{logical_gate}).
\item {\bf Connection to D-Wave machines and other Ising solvers}: As PyQUBO is included in the D-Wave Ocean package, PyQUBO can easily be integrated with the Ocean solver for D-Wave machines (Sec. \ref{dwave}). The format of QUBOs produced by PyQUBO is a key-value dictionary, which is compatible with other software tools for Ising solvers, such as OpenJij \cite{openjij}.
\item {\bf Validation of a QUBO created using PyQUBO}: The PyQUBO utility method \texttt{utils.asserts. assert\_qubo\_equal()} checks the equality of given QUBOs, which are symmetrized so that they can be compared.
\end{itemize}
}

\leftskip=0pt

\section{PyQUBO Fundamentals}
\label{sec:pyqubo_fundamentals}

In this section, we will explain the essential classes of PyQUBO required to write and solve basic combinatorial optimization problems. 

PyQUBO can be installed using \texttt{pip} as follows:
\begin{lstlisting}[language=bash]
pip install pyqubo
\end{lstlisting}
Alternatively, GitHub users can install it from the source code:
\begin{lstlisting}[language=bash]
git clone https://github.com/recruit-communications/pyqubo.git
cd pyqubo
python setup.py install
\end{lstlisting}
Supported Python versions are listed on the Github repository page.

\subsection{Defining the Hamiltonian with the Express Class}

 The \texttt{Express} class is the abstract class of all operations used to write Hamiltonians in PyQUBO. Once defined, the Hamiltonian can easily be converted into a binary quadratic problem using the \texttt{compile()} class method. 

\subsubsection{Spin and Binary}
\label{spin-binary}

The Hamiltonians of combinatorial optimization problems can be expressed in terms of the \texttt{Spin} $(-1, 1)$ and \texttt{Binary} $(0, 1)$ classes, corresponding to the Ising model and QUBO formulations, respectively. A \texttt{Binary} or \texttt{Spin} that is assigned to a variable must also be provided with a unique label. The labels of \texttt{Binary}/\texttt{Spin} variables can be used to interpret the coefficients of a QUBO or Ising problem more efficiently.
% Example with both binary and spin, and corresponding mathematical equations.
\begin{lstlisting}[language=Python, caption={Creating spins and binaries in PyQUBO.}]
>>> from pyqubo import Binary, Spin
>>> a, b, c = Binary("a"), Binary("b"), Binary("c")
>>> p, q, r = Spin("p"), Spin("q"), Spin("r")
\end{lstlisting}

\subsubsection{Add, Mul, and Num}

PyQUBO interprets the built-in addition, multiplication, and power operators of Python, as well as the Python \texttt{int} and \texttt{float} values, as \texttt{Express} instances. \texttt{Spin} and \texttt{Binary} can also be added or multiplied using the \texttt{Add} and \texttt{Mul} classes, whereas numerical constants can be written using the \texttt{Num} class, given a float or integer as its parameter.

The following example is a simple Hamiltonian that is minimized when one of the binary variables $a$ or $b$ is $1$:
\begin{equation}
\label{eq:simple}
    H = (a \times b - 1)^{2},\quad a,b\in \{0,1\}.
\end{equation}
Codeblock \ref{lst:add} demonstrates how we can express the Hamiltonian using PyQUBO:
\begin{lstlisting}[language=Python, label={lst:add}, caption={Arithmetic of \texttt{Binary} or \texttt{Spin} express variables is made possible using Python operators.}]
>>> from pyqubo import Binary
>>> H = (Binary("a") * Binary("b") - 1)**2
\end{lstlisting}
\vspace{-2mm}
\subsection{Compilation of Express Instances}
\label{model-and-solver}

\subsubsection{From Expression to Model}
% Reduction of the degree can be separated into another section.
% Mathematical explanation in previous paper
A \texttt{Model} instance is created from an \texttt{Express} class using the \texttt{compile()} method. \texttt{Model} contains information regarding the Hamiltonian of QUBO and Ising representations. Codeblock \ref{lst:compile} shows how we compile Eq. (\ref{eq:simple}) from the previous section.
\begin{lstlisting}[language=Python, label={lst:compile}, caption={Creating a model from a PyQUBO expression.}]
>>> from pyqubo import Binary
>>> H = (Binary("a") * Binary("b") - 1)**2
>>> model = H.compile()
\end{lstlisting}
\vspace{-2mm}
\subsubsection{From Model to QUBO or Ising}
A model's QUBO or Ising formulations can be retrieved as Python dictionaries using the model class' \texttt{to\_qubo()} and \texttt{to\_ising()} methods. The \texttt{to\_qubo()} method returns the QUBO and its energy offset. The QUBO takes the form of \texttt{dict[(label, label), value]}, where each label corresponds to a variable. The \texttt{to\_ising()} method returns the Ising model as two dictionaries, corresponding to linear and quadratic terms, as well as their energy offset. The linear and quadratic outputs take the form \texttt{dict[label, value]} and \texttt{dict[(label, label), value]}, respectively. Codeblock \ref{lst:toqubo} shows Eq. (\ref{eq:simple}) represented as both a QUBO and an Ising model with energy offsets using the model constructed in Codeblock \ref{lst:compile}.
\begin{lstlisting}[language=Python, firstnumber=4, label={lst:toqubo}, caption={Converting a model into a QUBO or Ising model returns an energy offset as well.}]
>>> qubo, qubo_offset = model.to_qubo()
>>> qubo |\Suppressnumber|
{('a', 'b'): -1.0, ('b', 'b'): 0, ('a', 'a'): 0} |\Reactivatenumber|
>>> qubo_offset |\Suppressnumber|
1.0 |\Reactivatenumber|
>>> linear, quad, ising_offset = model.to_ising()
>>> linear, quad |\Suppressnumber|
({'b': -0.25, 'a': -0.25}, {('a', 'b'): -0.25}) |\Reactivatenumber|
>>> ising_offset |\Suppressnumber|
0.75 |\Reactivatenumber|
\end{lstlisting}

\subsubsection{Order Reduction through Compilation}
\label{sec:order_reduction}

During compilation, if a Hamiltonian includes $k$-body interactions among \texttt{Binary} or \texttt{Spin} variables for $k > 2$, PyQUBO automatically reduces the expression to a quadratic by creating auxiliary variables representing the products of individual spins or binaries. 

Reducing the order of the Hamiltonian by hand can be complicated, even for $k = 3$ or three-body interactions. For example, given the Hamiltonian $H = xyz$, where $x$, $y$, and $z$ are binary variables, we introduce the auxiliary binary $a$, which represents the product of $x$ and $y$. However, to maintain the relationship $a = xy$, we must also include the penalty term $AND(a,x,y) \equiv xy - 2a(x+y) + 3a$ to the Hamiltonian. The final Hamiltonian is $H = az + \alpha AND(a, x, y)$, where $\alpha$ is the penalty strength. 

Meanwhile, PyQUBO performs this reduction automatically, as demonstrated in Codeblock \ref{lst:order-reduction}. The QUBO created in line 5 introduces a new variable labeled \texttt{'x*y'} representing the product of the individual binary variables.
\begin{lstlisting}[language=Python, label={lst:order-reduction}, caption={\texttt{Model.compile()} reduces the degree of an expression if it is greater than two.}]
>>> from pyqubo import Binary
>>> x, y, z = Binary("x"), Binary("y"), Binary("z")
>>> alpha = 2.0
>>> model = (x*y*z).compile(strength=alpha)
>>> qubo, offset = model.to_qubo()
>>> print(qubo) |\Suppressnumber|
{('x', 'y'): 5.0, ('x', 'x*y'): -10.0, 
 ('x*y', 'y'): -10.0, ('x*y', 'z'): 1.0, 
 ('x', 'x'): 0.0, ('y', 'y'): 0.0, 
 ('x*y', 'x*y'): 15.0, ('z', 'z'): 0} |\Reactivatenumber|
\end{lstlisting}

\subsubsection{Decode Solutions}
\label{decode_solution}
The \KTadd{\texttt{Model.decode\_sample()} method} can interpret the solution from any PyQUBO or quantum annealing solver as an easy-to-read Python dictionary of variable labels and their corresponding values ($0$ or $1$ for \texttt{vartype="BINARY"}; $-1$ or $1$ for \texttt{vartype="SPIN"}). \KTadd{The function returns a \texttt{DecodedSample} object, which provides a dictionary of label-value pairs via the \texttt{sample} property, the energies of constraints via the \texttt{constraints()} method, and the energy of the solution. We will discuss the \texttt{constraints()} method in greater detail in Section \ref{subsec:constraint}.}
\begin{lstlisting}[language=Python, label={lst:decode}, caption={Decoding a solution.}]
|\KTadd{>>> decoded\_sample}| = model.decode_sample(sol, vartype="BINARY")
|\KTadd{>>> decoded\_sample.sample}\Suppressnumber{}|
|\KTadd{\{'a': 1, 'b': 1\}}\Reactivatenumber{}|
|\KTadd{>>> decoded\_sample.constraints()}\Suppressnumber{}|
|\KTadd{\{'const1': (False, -3.0)\}}\Reactivatenumber{}|
|\KTadd{>>> decoded\_sample.energy}\Suppressnumber{}|
|\KTadd{-3.0}\Reactivatenumber{}|
\end{lstlisting}
\vspace{-2mm}
\subsection{The Number Partioning Problem}
\label{number-partitioning}
Using the classes described in Sections \ref{spin-binary} and \ref{model-and-solver}, we solve the \textbf{number partitioning problem}, which is described as follows. Given a set $S$ of $N$ positive integers, create two disjoint subsets of integers $S_{1}$ and $S_{2}$ such that their sums are equal. The Ising formulation of the number partitioning problem is
\begin{equation}
H = \left( \sum_{i = 1}^{N} n_{i}s_{i}\right )^2
\end{equation}
where \(n_{i}\) describes the numbers in the set \(S\) and \(s_{i}\) is a spin variable. Given that \(s_{i}\) takes the value $1$ or $-1$, the sum of two equally sized sets will be zero for optimal solutions.

Codeblock \ref{lst:spin} shows how the Hamiltonian of the number partitioning problem with $S=\{4,2,7,1\}$ can be prepared and solved using PyQUBO.
\begin{lstlisting}[language=Python, label={lst:spin}, caption={We use \texttt{Spin} variables to distinguish the two sets of integers, which are represented by the coefficient values.}]
|\KTadd{>>> import neal}|
>>> from pyqubo import Spin, solve_ising
>>> s1, s2, s3, s4 = Spin("s1"), Spin("s2"), Spin("s3"), Spin("s4")
>>> H = (4*s1 + 2*s2 + 7*s3 + s4)**2
>>> model = H.compile()
>>> qubo, offset = model.to_qubo()
|\KTadd{>>> sampler = neal.SimulatedAnnealingSampler()}|
|\KTadd{>>> sampleset = sampler.sample\_qubo(qubo)}|
|\KTadd{>>> decoded\_samples =model.decode\_sampleset(}\Suppressnumber{}|
      |\KTadd{sampleset, vartype="SPIN")}\Reactivatenumber{}|
|\KTadd{>>> best = min(decoded\_samples, key=lambda x: x.energy)}|
|\KTadd{>>> best.sample} \Suppressnumber{}|
|\KTadd{\{'s1': -1, 's2': -1, 's4': -1, 's3': 1\}} \Reactivatenumber{}|
\end{lstlisting}

% add the output of decoded object
% how to interpret the solution

Here the detail explanation of Codeblock \ref{lst:spin} is given as follows.
The sum of the multiplied terms and its exponent are represented using the corresponding Python addition ($+$), multiplication ($*$), and exponent ($**$) operators, as indicated in line \KTadd{4}.
In lines \KTadd{5 and 6}, the Hamiltonian is compiled and converted into a QUBO, and in line 8, a possible solution is identified using the \KTadd{\texttt{sample\_qubo()} method of \texttt{SimulatedAnnealingSampler}. In line 9, \texttt{decode\_sampleset()}} is used to retrieve the spin values corresponding to the labels used in line \KTadd{3}. The solution can easily be validated by comparing the sums of the coefficients corresponding to the spins with the value 1 or $-1$. The solution demonstrates one manner in which the given set of integers can be separated into subsets such that their sums are equal. That is, other solutions where the difference is $0$ may exist, whereas in other cases, depending on the set of integers provided, none may exist at all. 

\section{PyQUBO Advanced Use}
\label{advanced_usage}
Numerous combinatorial optimization problems can be written and solved using the PyQUBO classes explained in Section 6. Advanced users interested in representing larger or more complicated problems can take advantage of PyQUBO's many other utilities or even define their own Hamiltonian class using the \texttt{UserDefinedExpress} class.
In this section, we explain the \texttt{Constraint}, \texttt{Placeholder}, and \texttt{Integer} classes of PyQUBO and demonstrate the manner in which these can be used to write more complicated combinatorial optimization problems. 
\vspace{-2mm}
\subsection{Constraint}
\label{subsec:constraint}

In Section 6.3, we observed that a solution to the number partitioning problem, given a small set $S$, is easily validated by hand. However, this process becomes difficult and time consuming for larger problems with many auxiliary variables, such as the knapsack problem. The \texttt{Constraint} class provides automatic validation regarding whether a solution satisfies the given constraints. 
The \texttt{Constraint} class specifies the parts of a Hamiltonian that must be satisfied by a valid solution to an optimization problem. Each \texttt{Constraint} instance takes the section of the Hamiltonian comprising a constraint and string label as its parameters.
If a Hamiltonian defined with the \texttt{Constraint} class is compiled as a model, the \KTadd{\texttt{decode\_sample()} function returns a \texttt{DecodedSample} object that also provides information about the constraints via the \texttt{constraints()} method. The information is represented as a dictionary of constraint labels and tuples containing a boolean value and number, which correspond to whether the constraint is satisfied and the energy of the penalty term, respectively. The \texttt{DecodedSample} object also contains the energy of a given solution via the \texttt{energy} property, as well as corresponding variable and value pairs in the \texttt{sample} property.}
Codeblock \ref{lst:constraint} shows a simple application of the \texttt{Constraint} class. In line 3, we create a sum of two binary variables with a penalty term that is minimized when one of the variables is $1$ and the other is $0$. Because the penalty term is wrapped with the \texttt{Constraint} class, we are able to use the \KTadd{\texttt{decode\_sample()}} method to check whether a given solution satisfies the constraint represented by the penalty term. Lines 5 to 8 show that when both binary values are $1$, \KTadd{\text{constraints()} method returns a false value with a label \texttt{one\_hot} indicating the constraint is not satisfied. When the feasible solution ($a=1, b=0$) is given, the \text{constraints()} method returns a true value indicating the constraint is satisfied.}
\begin{lstlisting}[language=Python, label={lst:constraint}, caption={When used with the \texttt{Constraint} class, \KTadd{\texttt{decode\_sample()} will return information about the constraints.}}]
>>> from pyqubo import Binary, Constraint
>>> a, b = Binary('a'), Binary('b')
>>> exp = a+b+Constraint((a+b-1)**2, label='one_hot')
>>> model = exp.compile()
|\KTadd{>>> decoded\_sample = model.decode\_sample(}\Suppressnumber{}|
|\KTadd{~~~~\{'a': 1, 'b': 1\}, vartype='BINARY')}\Reactivatenumber{}|
|\KTadd{>>> print(decoded\_sample.constraints())}\Suppressnumber{}|
|\KTadd{\{'one\_hot': (False, 1.0)\}}\Reactivatenumber{}|
|\KTadd{>>> decoded\_sample = model.decode\_sample(}\Suppressnumber{}|
|\KTadd{~~~~\{'a': 1, 'b': 0\}, vartype='BINARY')}\Reactivatenumber{}|
|\KTadd{>>> print(decoded\_sample.constraints())}\Suppressnumber{}|
|\KTadd{\{'one\_hot': (True, 0.0)\}}\Reactivatenumber{}|
\end{lstlisting}
\vspace{-2mm}
\subsection{Placeholder}
\label{subsec:placeholder}

Depending on the given Hamiltonian, the compilation of models can be computationally expensive. If any value in the Hamiltonian is changed, the model must be recompiled in order to calculate the new QUBO. The \texttt{Placeholder} class makes it possible to update the constants and coefficients within the Hamiltonian without recompiling the model, thereby saving a significant amount of time in the situation where we need to update some values in Hamiltonian, such as parameter tuning of the penalty strength. Placeholders substitute constants and are identified by their string labels. When creating a QUBO or Ising models, or when decoding a solution, \texttt{Placeholder} values must be specified using a Python dictionary of label-value pairs as an additional \texttt{feed\_dict} parameter.

Codeblock \ref{lst:placeholder} shows how the \texttt{Placeholder} can be used after compiling an expression into a \texttt{Model} instance. In lines 5 and 7, when converting the \texttt{Model} into a QUBO, we include a \texttt{feed\_dict} specifying the value of \texttt{Placeholder} \texttt{a}.
\begin{lstlisting}[language=Python, caption={\texttt{Placeholder} values can be updated after compiling a model.}, label={lst:placeholder}]
>>> from pyqubo import Binary, Placeholder
>>> x, y, a = Binary('x'), Binary('y'), Placeholder('a')
>>> exp = a*x*y + 2.0*x
>>> model = exp.compile()
>>> qubo_1 = model.to_qubo(feed_dict={'a': 3.0})
>>> qubo_1 |\Suppressnumber{}|
({('x', 'x'): 2.0, ('x', 'y'): 3.0, ('y', 'y'): 0}, 0.0) |\Reactivatenumber{}|
>>> qubo_2 = model.to_qubo(feed_dict={'a': 5.0}) 
>>> qubo_2 |\Suppressnumber{}|
({('x', 'x'): 2.0, ('x', 'y'): 5.0, ('y', 'y'): 0}, 0.0) |\Reactivatenumber{}|
\end{lstlisting}

\vspace{-3mm}
\subsection{Integer} 
\label{subsec:integer}

The \texttt{Integer} class can easily create integer encodings of variables using the sums of \texttt{Binary} terms with various encoding types. PyQUBO supports four types of integer encoding classes: \texttt{OneHotEncInteger}, \texttt{UnaryEncInteger}, \texttt{LogEncInteger} and \texttt{OrderEncInteger} (Supplementary Section B). All four classes require a label, \KTadd{a tuple of} lower bound, and upper bound as arguments, and represent values in the range \([lower, upper]\), where $lower$ and $upper$ are the lower and upper bound value, respectively.

\subsection{The Knapsack Problem}
\label{knapsack}

We use the PyQUBO classes described above to create and solve the \textbf{knapsack problem}, which is described as follows. Given a set of $N$ items with integer weights and values, determine which items to include in a collection such that the total weight is less than or equal to a weight limit $W$ and the total value $V$ is as large as possible.
The total weight and total value of the knapsack is represented by
\begin{align}
    W &= \sum_{\alpha=1}^{N} w_{\alpha} x_{\alpha},\quad w_{\alpha} \in \mathcal{Z}\\
    V &= \sum_{\alpha=1}^{N} v_{\alpha} x_{\alpha},\quad v_{\alpha} \in \mathcal{Z},
\end{align}
where \(x_{\alpha}\) takes the value $0$ or $1$ depending on whether the object \(\alpha\) is in the knapsack, and \(w_{\alpha}\) and \(v_{\alpha}\) are the integer weight and value of each item \(\alpha\), respectively. 
The Hamiltonian of the knapsack problem takes the form \(H = H_{A} - H_{B}\), where \(H_{A}\) represents the weight constraint and \(H_{B}\) is the total value of the items collected. Here, since Ising machines solve minimization problem, the sign of the second term is set to minus. The definitions of \(H_{A}\) and \(H_{B}\) are as follows:
\begin{equation}
\label{eq:knapsack_HA}
H_{A} = \lambda_{1}\left(1-\sum_{n = 1}^{W} y_{n} \right)^2
+ \lambda_{2}\left (\sum_{n = 1}^{W} n y_{n} - \sum_{\alpha = 1}^{N} w_{\alpha} x_{\alpha} \right)^2,
\end{equation}
\begin{equation}
\label{eq:knapsack_HB}
H_{B} = \sum_{\alpha=1}^{N} v_{\alpha} x_{\alpha},
\end{equation}
where \(y_{n}\) takes either the value $1$ if the knapsack weight is \(n\) or $0$ otherwise \cite{lucas2014ising, tanahashi2019application}. 

Below, we demonstrate how the knapsack problem can be written and solved using PyQUBO.
% Include for loop iteration.
\begin{lstlisting}[language=Python, basicstyle=\ttfamily\scriptsize\color{black},rulecolor=\color{black},numberstyle=\color{black}, label={lst:knapsack}, caption = {While the knapsack problem can be expressed and solved using the classes introduced in Section \ref{sec:pyqubo_fundamentals} alone, the \texttt{Constraint}, \texttt{Placeholder}, \texttt{Array}, and \texttt{Integer} classes are useful for both debugging and streamlining the code.}]
from pyqubo import Binary, Constraint, Placeholder, Array, OneHotEncInteger

weights = [1, 3, 7, 9]
values = [10, 2, 3, 6]
max_weight = 10

# create the array of 0-1 binary variables
# representing the selection of the items
n=len(values)
items = Array.create('item', shape=n, vartype="BINARY")

# define the sum of weights and values using variables
knapsack_weight = sum(|\Suppressnumber{}|
    weights[i] * items[i] for i in range(n))|\Reactivatenumber{}|
knapsack_value = sum(|\Suppressnumber{}|
    values[i] * items[i] for i in range(n))|\Reactivatenumber{}|

# define the coefficients of the penalty terms,
# lmd1 and lmd2, using Placeholder class
# so that we can change their values after compilation
lmd1 = Placeholder("lmd1")
lmd2 = Placeholder("lmd2")

# create Hamiltonian and model
weight_one_hot = OneHotEncInteger("weight_one_hot", value_range=(1, max_weight), strength=lmd1)
Ha = Constraint((weight_one_hot - knapsack_weight)**2, "weight_constraint")
Hb = knapsack_value
H = lmd2*Ha - Hb
model = H.compile()

# use simulated annealing (SA) sampler of neal package
sampler = neal.SimulatedAnnealingSampler()

feasible_sols = []
# search the best parameters: lmd1 and lmd2
for lmd1_value in range(1, 10):
    for lmd2_value in range(1, 10):
        
        feed_dict = {'lmd1': lmd1_value, "lmd2": lmd2_value}
        qubo, offset = model.to_qubo(feed_dict=feed_dict)
        bqm = model.to_bqm(feed_dict=feed_dict)
        bqm.normalize()
        sampleset = sampler.sample(bqm, num_reads=10, sweeps=1000, beta_range=(1.0, 50.0))
        dec_samples = model.decode_sampleset(sampleset, feed_dict=feed_dict)
        best = min(dec_samples, key=lambda x: x.energy)
        
        # store the feasible solution
        if not best.constraints(only_broken=True):
            feasible_sols.append(best)

best_feasible = min(feasible_sols, key=lambda x: x.energy)
print(f"selection = {[best_feasible.sample[f'item[{i}]'] for i in range(n)]}")
print(f"sum of the values = {-best_feasible.energy}")|\Suppressnumber{}|

[output]
selection = [1, 0, 0, 1]
sum of the values = 16.0|\Reactivatenumber{}|
\end{lstlisting}

The detail of Codeblock \ref{lst:knapsack} is as follows. In lines \KTadd{3-5}, we define the weights and values of the items in our set, as well as our weight limit $W$. In line \KTadd{10}, we use the \texttt{Array} class to create the set of \texttt{Binary} variables \(x_{\alpha}\), with its size set equal to the number of items. Thereafter, we calculate the knapsack weight and value sums $\sum_{\alpha=1}^{N} w_{\alpha}x_{\alpha}$ and $\sum_{\alpha=1}^{N} v_{\alpha}x_{\alpha}$ in lines \KTadd{13 and 14}, respectively. 

Next, we prepare the Hamiltonian. In lines \KTadd{19-20}, we define \texttt{Placeholder} instances representing the coefficient \KTadd{$\lambda_{1}$ and $\lambda_{2}$} in Eq. (\ref{eq:knapsack_HA}). In line \KTadd{23}, we create a \texttt{OneHotEncInteger}, with $\lambda_{1}$ as its \texttt{strength} argument, to represent the integer value $\sum_{n=1}^W ny_{n}$ for the range given by the weight limit (Eq. (\ref{eq:knapsack_HA})). In line \KTadd{24}, we take the squared difference between the \texttt{Integer} instance and knapsack weight sum, and wrap it with the \texttt{Constraint} class. In line \KTadd{25}, $H_{B}$ is written as the knapsack value sum.

To solve our Hamiltonian, we first compile it into a model instance in line \KTadd{27}. Then, we initialize the \texttt{Placeholder} value as a Python dictionary. In lines \KTadd{34-47}, we create a loop in which we search the values of $\lambda_{1}$ and $\lambda_{2}$ so that the feasible solutions can be obtained. In the loop, we create a QUBO as \KTadd{a \texttt{BQM} instance using \texttt{to\_bqm}}, normalize the QUBO matrix of \texttt{bqm} using \texttt{normalize} in order to make it easy to tune the solver parameters, sample solutions through simulated annealing using the \KTadd{\texttt{SimulatedAnnealingSampler} of the neal package, and then retrieve the list of \texttt{DecodedSample} instances using \texttt{decode\_sampleset()}. 
The \texttt{BQM} class is defined in the D-Wave's \texttt{dimod} package and represents QUBOs or Ising models.
The method decodes a \texttt{SampleSet} object and returns a \texttt{DecodedSample} like \texttt{decode\_sample()} method. In lines 46-47, we retrieve the broken constraints using \texttt{constraints} with the \texttt{only\_broken} argument set to true.
If the solution is feasible, we append the solution to the list \texttt{feasible\_sols}. In line 49, we obtain the feasible solution with the lowest energy. We obtained the final solution to select 1st and 4th items with the value sum equal to $16$.}

\vspace{-3mm}
\section{PyQUBO for Logical Gates}
\label{logical_gate}

A logical gate is an electronic device that implements a boolean function by taking binary inputs, performing a logic operation, and providing a single binary output. 

The AND, OR, NOT, and XOR logic operations can be represented using the logical gate and logical constraint classes of PyQUBO. In the following, we present the corresponding circuit component and truth table of each logic class, as well as examples of its application in PyQUBO. 
\vspace{-2mm}
\subsection{Logical Gates}
\label{subsec:logical_gate}

The PyQUBO logical gate classes \texttt{Not}, \texttt{And}, \texttt{Or}, and \texttt{Xor} correspond to the four logic operations. Like their electronic analogs, the logical gate classes perform the specified logic operation on given binary inputs. 

\begin{figure}[h]
    \centering
    \includegraphics[width=9cm]{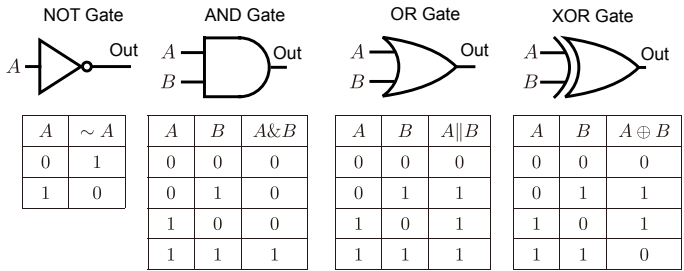}
    \caption{A summary of logical gate circuit elements and their truth tables.}
    \label{fig:circuits}
\end{figure}

We summarize the analytical representations of the logic operations below:
\begin{align}
    NOT(a) &= 1 - a \\
    AND(a,b) &= ab \\
    OR(a,b) &= a + b - ab \\
    XOR(a,b) &= a + b - 2ab,
\end{align}
where $a$ and $b$ represent binary inputs.
Figure \ref{fig:circuits} shows each operation's circuit element and truth table.

\vspace{-3mm}
\subsection{Logical Constraints}
\label{subsec:logical_constraints}

The logical constraint classes \texttt{NotConst}, \texttt{AndConst}, \texttt{OrConst}, and \texttt{XorConst} are constraints based on the four operators discussed. However, unlike the logical gates in Section \ref{subsec:logical_gate}, each constraint requires three \texttt{Express} class inputs (or two in the case of \texttt{NotConst}) corresponding to the logical expression operands, as well as the expected output and a label. In other words, the logical constraint classes, instead of performing the specified operation, represents an entire logic expression as a constraint.

As discussed in Section \ref{decode_solution}, the \texttt{decode\_sample()} function determines whether a given solution violates the constraints. Another approach to verifying the solution of an expression wrapped by a logical constraint is to determine the model energy, as indicated in Codeblock \ref{lst:orconst} using \texttt{OrConst}. When the provided variables satisfy the constraint, the energy is $0.0$, but when they break the constraint, the energy is $1.0$. Therefore, unlike the logical gate classes, logical constraints do not provide solutions themselves.
\begin{lstlisting}[language=Python, label={lst:orconst}, caption={The energy of a compiled logical constraint expression indicates whether or not the provided operands and output satisfy the operator logic. In this example, the energy is $0.0$ when the expression is true and $1.0$ when it is false.}]
>>> from pyqubo import OrConst, Binary
>>> a, b, c = Binary('a'), Binary('b'), Binary('c')
>>> exp = OrConst(a, b, c, 'or')
>>> model = exp.compile()
>>> model.energy({'a': 1, 'b': 0, 'c': 1}, vartype='BINARY') |\Suppressnumber{}|
0.0 |\Reactivatenumber{}|
>>> model.energy({'a': 0, 'b': 1, 'c': 0}, vartype='BINARY') |\Suppressnumber{}|
1.0 |\Reactivatenumber{}|
\end{lstlisting}
\vspace{-3mm}
\subsubsection{Preparing a Multi-bit Binary Multiplier}
\label{subsec:binary_multiplier}

\begin{figure}[htp]
    \centering
    \includegraphics[width=8cm]{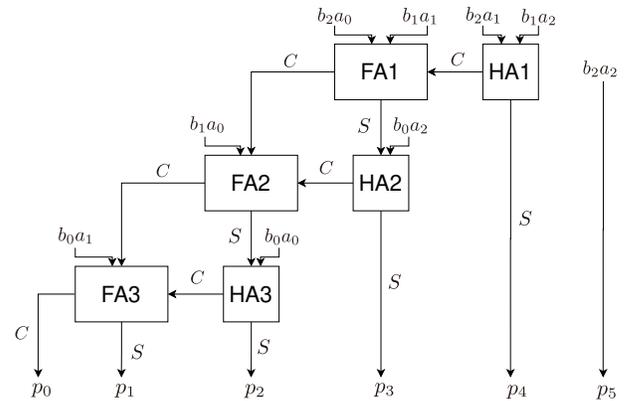}
    \caption{Diagram for a 3-bit binary multiplier. The arrows labeled \(C\) and \(S\) refer to the sum and carry of the individual half- and full-adders (labeled HA and FA). The \(b_{m}a_{n}\) labels represent the $m$-th and $n$-th digit binaries joined by an AND gate.}
    \label{fig:multiplier-circuit}
\end{figure}

\begin{figure}[!htp]
\centering
\includegraphics[width=5.5cm]{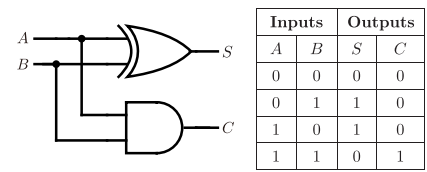}
\caption{Half-adder circuit and truth table.}
\label{fig:half-adder}
\end{figure}

A digital binary multiplier is an electronic circuit that is used to multiply two binary numbers, namely, a multiplier and multiplicand. The bit size of the resulting product corresponds to the sum of the two input bit sizes. Digital multipliers consist of AND gates as well as half- and full-adders. For a $j$-bit multiplier and $k$-bit multiplicand, a multiplier circuit requires \(j \times k\) AND gates and \((j-1) \times k\) adders, as illustrated in Fig. \ref{fig:multiplier-circuit}.
The logical gates of PyQUBO can be used to represent the AND, OR, and XOR gates that comprise the multiplier as well as its half- and full-adders. By using logical constraints instead of logical gates, the binary multiplier can be adapted to solve integer factoring problems. For example, given a fixed product \textbf{p} as a constraint, the multiplier QUBO can be solved for all integer pairs \textbf{a} and \textbf{b} satisfying the product \cite{maezawa2017design}.
To prepare the digital multiplier, first we need to construct the half-adder and full-adder constraints.
The half-adder adds two single binary digits and returns their sum and carry values as shown in the circuit diagram in Fig. \ref{fig:half-adder}.

Codeblock \ref{lst:half-adder} demonstrates how a half-adder is built using the PyQUBO logical constraint classes. 

\begin{lstlisting}[language=Python, label={lst:half-adder}, caption = {Half-adder built using PyQUBO's logical constraint classes. Note that the outputs \(s\) and \(c\) must also be included as arguments.}]
from pyqubo import XorConst, AndConst, OrConst
from pyqubo import Binary, Spin, UserDefinedExpress
import dimod

class HalfAdderConst(UserDefinedExpress):
    def __init__(self, a, b, s, c, label):
        self.xor_const = XorConst(a,b,s,f'{label}_xor')  
        self.and_const = AndConst(a,b,c,f'{label}_and')
        express = self.and_const + self.xor_const
        super().__init__(express)

a, b, s, c = Binary("a"), Binary("b"), Binary("s"), Binary("c")
model = HalfAdderConst(a, b, s, c, 'ha').compile()
qubo, offset = model.to_qubo()

samplset = dimod.ExactSolver().sample_qubo(qubo)
|\KTadd{for s in model.decode\_sampleset(samplset):}|
    |\KTadd{print(s.sample['a'], s.sample['b'], s.sample['s'],}\Suppressnumber{}|
      |\KTadd{s.sample['c'], s.energy)}|

# Output
# 0 0 0 0 0.0
# 0 1 1 0 0.0
# 1 0 1 0 0.0
# 1 1 0 1 0.0
# 0 1 1 1 1.0
# 1 1 1 1 1.0
# ...|\Reactivatenumber{}|
\end{lstlisting}

\begin{figure}[!htp]
    \centering
    \includegraphics[width=8.5cm]{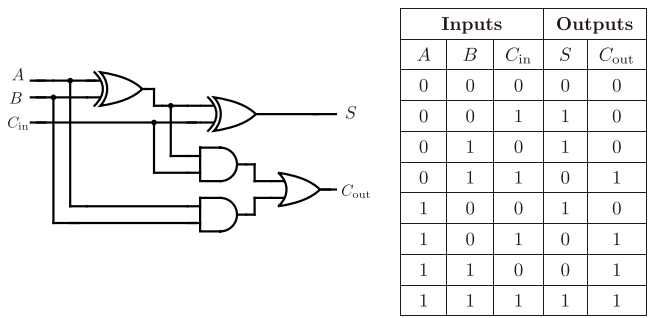}
    \caption{Full-adder circuit diagram and truth table, where $A$, $B$, and $C_{\rm in}$ are binary inputs.}
    \label{fig:full-adder}
\end{figure}

The detail of Codeblock \ref{lst:half-adder} is as follows. We create a new class \texttt{HalfAdderConst} that creates the Hamiltonian for a half-adder given two inputs, as well as the sum and carry. In lines 7 and 8, we define the XOR and AND constraints, which provide the sum and carry, respectively. In line 9, we define the Hamiltonian as the sum of the two constraints.
Line 16 uses the \texttt{dimod.ExactSolver()} method to calculate all possible solutions of the half-adder Hamiltonian. As seen in Codeblock \ref{lst:orconst}, valid solutions have $0$ energy.
% Using the logic constraint classes requires two additional input variables to designate the half-adder sum and carry. Solving the resultant QUBO enables us to determine valid half-adder outputs without iterating the energy function. For example, if we define the sum or carry variables as a fixed value, we can solve the QUBO to determine all operand combinations satisfying the new constraint. This approach can be scaled for solving integer factorization problems, as demonstrated in the following section.
The full-adder takes three binary values -- the two operands \(A\) and \(B\) and a carry value \(C_{\rm in}\) from another adder -- and outputs the sum \(S\) and \(C_{\rm out}\), as with the half-adder.
As indicated in Fig. \ref{fig:full-adder}, the full-adder can be constructed from two half-adders and an OR gate. Here, \(A\) and \(B\) are the inputs of one half-adder, whereas their sum-output and \(C_{\rm in}\) are the inputs of the second half-adder. The two half-adder carries are joined by the OR gate to output \(C_{\rm out}\), and the sum-output of the second half-adder provides \(S\).

A simple 3-bit binary multiplier uses three half-adders, three full-adders, and nine AND constraints. It takes the multiplicand, multiplier, and product as lists or arrays of binary values (as well as a unique string label as a prefix) and creates a Hamiltonian. Users interested in designing a binary multiplier using PyQUBO should refer to the PyQUBO GitHub repository\cite{pyqubo_github}.

\vspace{-2mm}
\section{PyQUBO with D-Wave Ocean}
\label{dwave}

In this section, we demonstrate how PyQUBO and D-Wave can be used in tandem to solve the knapsack problem once again (Eqs. (\ref{eq:knapsack_HA}) and (\ref{eq:knapsack_HB})). 

\vspace{-2mm}
\subsection{Introduction to D-Wave}
To use a D-Wave machine to solve a given problem, the logical graph representing the corresponding QUBO or Ising model must be embedded into the physical graph of D-Wave hardware. D-Wave machines use the Chimera or Pegasus architecture, depending on the hardware generation \cite{albash2015consistency, dwave:hardware-topology}. Here, qubit is a spin variable in quantum annealing machines.
%These unit cells are tiled as a lattice of sparsely connected qubits.
Embedding a model into the target architecture requires grouping multiple qubits into \textbf{chains} that represent single theoretical qubits \cite{choi2008minor,cai2014practical,boothby2016fast}. In addition to the problem QUBO, chained qubits are assigned the interaction with a constant \textbf{chain strength} so that their values are the same across all low-energy solutions. While the magnitude of the chain strength must be tuned depending on the problem, the embedding process allows even complex structures to be mapped onto the D-Wave machine.   

D-Wave System solvers are configured to solve problems on corresponding working graphs, which are the qubits and couplers that are available for computation. Here, couplers adjusts the value of the interaction.
% However, the yield of a working graph is usually less than the total number of qubits and couplers that is physically available owing to manufacturing and environmental variations \cite{albash2015consistency}. A virtual full-yield chimera (VFYC) solver enables developers to determine variables that cannot be represented by the working graph through hybrid use of the system QPU and post-processing \cite{dwave:vfyc}. 

\subsection{Programming with PyQUBO and D-Wave Ocean}

\texttt{Model} written in PyQUBO can be used as inputs for the D-Wave Ocean sampler. Samplers provide a sample set of solutions from the low-energy states of the objective function of an optimization problem. Creating a D-Wave sampler requires an endpoint, a D-Wave Application Programming Interface (API) token, which is available to all D-Wave Leap accounts, and a specific solver name. 
Based on Codeblock \ref{lst:dwave-solver}, a method for solving a knapsack problem using the D-Wave sampler is specifically described. \KTadd{In lines 13-24, we define the Hamiltonian of the knapsack problem. Here, we used \texttt{LogEncInteger} instead of \texttt{OneHotEncInteger} used in Codeblock \ref{lst:knapsack} to show the integer class object can be easily replaced.}
In line \KTadd{26}, we create our sampler using the default API endpoint URL, our account API token, and the solver \KTadd{\textbf{Advantage\_system1.1}}. In line \KTadd{38}, we create our embedding, which we map onto the quantum annealing machine using the \texttt{FixedEmbeddingComposite} class in line \KTadd{40}. Defining our embedding before the QUBO saves time, however finding the optimal embedding could in turn lead to more efficient problem solving.
In lines \KTadd{42-49}, we define the sampler's keyword arguments, which depend on the selected D-Wave solver. The following parameters are common across all hardware solvers: \texttt{num\_reads} and \texttt{annealing\_time}, which correspond to the number of anneals and time per anneal respectively; \texttt{num\_spin\_reversal\_transforms}, which sets the number of gauge transformations to be performed on the problem \cite{boixo2014evidence}; and \texttt{auto\_scale}, which indicates whether $h_{i}$ and $J_{ij}$ of the Ising model (Eq. (\ref{eq:hamiltonian-graph})) are rescaled. 

The two parameters \texttt{chain\_strength} and \texttt{chain\_break\_fraction} in lines belong to the \texttt{FixedEmbeddingComposite} class, which maps problems to the sampler using the given embedding. As discussed above, the \texttt{chain\_strength} parameter must be tuned for the problem. 

Next, we create an objective function, which provides the solution to a QUBO, its energy, as well as broken constraints, given a feed dictionary for \texttt{Placeholder} values and the above-mentioned keyword arguments. In line 52, we create a QUBO \KTadd{as \texttt{BQM}} from a \texttt{Model} instance. In line \KTadd{53}, we normalize the QUBO \KTadd{such that the parameter tuning is not affected by the scale of the problem}. In line \KTadd{54}, we use the sampler to retrieve a set of solutions to the normalized QUBO, and in line 55, we use the PyQUBO function \KTadd{\texttt{Model.decode\_sampleset()}} to interpret these solutions. The decode method's arguments are \texttt{sampleset}, the solution from a sampler; and \texttt{feed\_dict}, which, as before, is a dictionary of placeholder key-value pairs. \KTadd{In line 56, we return the \texttt{DecodedSample} object with the lowest energy.}

\begin{figure}[!htp]
\centering
\includegraphics[width=7.5cm]{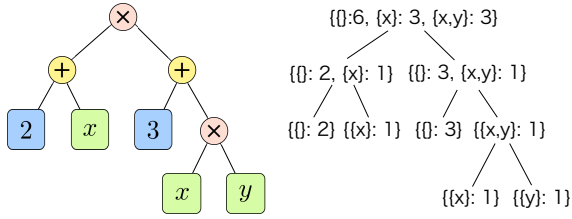}
\caption{\KTadd{(Left) The binary tree of the expression defined in the Codeblock \ref{lst:hamiltonian}. (Right) The generated polynomials at each node. The polynomials are shown as a hash map.}}
\label{fig:expression_bin_tree}
\end{figure}

Finally, in lines \KTadd{59-64}, we execute the objective function using different \texttt{Placeholder} values \KTadd{and append each feasible solution to the list \texttt{feasible\_sols}. In this code, we search only one parameter \texttt{lmd} since we are using \texttt{LogEncInteger}, which does not have an extra constraint like \texttt{OneHotEncInteger}.
Finally, we show the sum of the values of the best feasible solution.}
\begin{lstlisting}[language=Python,basicstyle=\ttfamily\scriptsize\color{black},rulecolor=\color{black},numberstyle=\color{black},label={lst:dwave-solver}, caption={Using the D-Wave sampler to find a solution to the knapsack problem.}]
import dimod
from dwave.system.samplers import DWaveSampler
from dwave.system.composites import FixedEmbeddingComposite
from minorminer.busclique import find_clique_embedding
import dwave_networkx as dnx
from pyqubo import Binary, Constraint, Placeholder, Array, LogEncInteger

# weights, values and the maximum weight of the knapsack problem
weights = [1, 3, 7, 9]
values = [10, 2, 3, 6]
max_weight = 10

n=len(values)
items = Array.create('item', shape=n, vartype="BINARY")
knapsack_weight = sum(weights[i] * items[i] for i in range(n))
knapsack_value = sum(values[i] * items[i] for i in range(n))

# create Hamiltonian and model
weight_one_hot = LogEncInteger("weight_one_hot", value_range=(1, max_weight))
Ha = Constraint((weight_one_hot - knapsack_weight)**2, "weight_constraint")
Hb = knapsack_value
lmd = Placeholder("lmd")
H = lmd*Ha - Hb
model = H.compile()

dw_sampler = DWaveSampler(
    endpoint="https://cloud.dwavesys.com/sapi", 
    token="your-token",
    solver="Advantage_system1.1")

graph_size=16
sampler_size=len(model.variables)
p16_working_graph = dnx.pegasus_graph(
                        graph_size,
                        node_list=dw_sampler.nodelist,
                        edge_list=dw_sampler.edgelist)

embedding = find_clique_embedding(sampler_size, p16_working_graph)

sampler = FixedEmbeddingComposite(dw_sampler, embedding)

sampler_kwargs = {
    "num_reads": 100,
    "annealing_time": 20,
    "num_spin_reversal_transforms": 4,
    "auto_scale": True,
    "chain_strength": 2.0,
    "chain_break_fraction": True
}

def objective(feed_dict):
    bqm = model.to_bqm(index_label=True, feed_dict=feed_dict)
    bqm.normalize()
    sampleset = sampler.sample(bqm,  **sampler_kwargs)
    dec_samples = model.decode_sampleset(sampleset, feed_dict=feed_dict)
    return min(dec_samples, key=lambda x: x.energy)

# search best parameters lmd within [1,2,...,5]
feasible_sols = []
for lmd_value in range(1, 5):
    feed_dict = {'lmd': lmd_value}
    s = objective(feed_dict)
    if not s.constraints(only_broken=True):
        feasible_sols.append(s)

best_feasible = min(feasible_sols, key=lambda x: x.energy)
print(f"selection = {[best_feasible.sample[f'item[{i}]'] for i in range(n)]}")
print(f"sum of value = {-best_feasible.energy}")|\Suppressnumber{}|

[output]
selection = [1, 0, 0, 1]
sum of value = 16.0|\Reactivatenumber{}|
\end{lstlisting}
\vspace{-10mm}
\KTadd{\section{Implementation and Benchmarking}
In this section, we show how PyQUBO is implemented internally and benchmark the performance with different implementations including other packages.
}
\vspace{-2mm}
\KTadd{\subsection{Internal Representation of Expressions}
In PyQUBO, the expression of a Hamiltonian is represented by a binary tree, which is called AST (Abstract Syntax Tree). For example, the expression created by the Codeblock \ref{lst:hamiltonian} is represented by the binary tree shown in Fig. \ref{fig:expression_bin_tree} (left). The leaves of the tree are composed of ``number'' and ``variable'' nodes, shown as rectangles in Fig. \ref{fig:expression_bin_tree} (left). The internal nodes are composed of ``sum'' or ``product'' nodes, shown as circles in Fig. \ref{fig:expression_bin_tree} (left).
}
\vspace{-2mm}
\KTadd{\subsection{Compilation (QUBO Creation)}
We define compilation in PyQUBO as the process to produce quadratic polynomials from the expression of the Hamiltonian. The compilation is composed of the following two steps.
\begin{itemize}
    \item Step 1: Expanding an expression into a polynomial.
    \item Step 2: Reducing the order of the polynomial obtained in the step 1.
\end{itemize}
We can easily implement order reduction by just replacing the pair of variables with the new auxiliary variable, as explained in \ref{sec:order_reduction}. In the following sections, we mainly explain the expansion of the expression (i.e., step 1).}

\KTadd{In the compilation process, the quadratic polynomial with binary variables corresponding to the QUBO matrix is produced from the binary tree representing the expression. In PyQUBO, polynomials are represented by a ``hash map'' with products of binary variables as keys and coefficients as values. Products of variables can be represented by ``set'' since we only deal with 0-1 binary variables. For example, we can confirm $xy^2z$ is represented by the set $\{x,y,z\}$ by using the relationship $xy^2z=xyz, ~x,y,z\in \{0,1\}$.}

\KTadd{When we represent the set with elements $A,B$ as $\{A,B\}$ and the hash map with key $k$ and value $v$ as $\{k: v\}$, the polynomial $a+bx_0+cx_0x_1$ can be written as $\{\{\}:a,~\{x_0\}:b,~\{x_0,x_1\}:c\}$, where $a,b$ and $c$ are coefficients, $x_0$ and $x_1$ are 0-1 binary variables, and $\{\}$ is an empty set.
}
\KTadd{
The Python-like pseudo code to expand the expression is shown in Codeblock \ref{lst:expand}. If we pass the root node object of the binary tree to the function, the expanded polynomial is returned as a hash map. The functions \texttt{poly\_sum}, \texttt{poly\_prod} calculate the sum and the product of the input polynomials, respectively. We assume that the input node has the property \texttt{type}, which indicates the type of the node (i.e., number, variable, sum or product). The ``product'' or ``sum'' node has the properties \texttt{left} and \texttt{right}, each of which contains the child node corresponding to the inputs of the product or sum operation. The ``number'' node has the property \texttt{value} which contains the number itself. The generated polynomials by \texttt{expand()} function at each node, are shown in Fig. \ref{fig:expression_bin_tree} (right).}
\begin{lstlisting}[language=Python, label={lst:hamiltonian}, caption={\KTadd{An example expression created by PyQUBO objects. It corresponds to the binary tree in Fig. \ref{fig:expression_bin_tree} (left).}}]
x, y = Binary('x'), Binary('y')
H = (x + 2)*(x*y + 3)
\end{lstlisting}

\vspace{-3mm}
\begin{lstlisting}[language=Python, label={lst:expand}, caption={\KTadd{The Python-like pseudo code to expand the expression.}}]
def expand(node):
  if node.type=='sum':
    return poly_sum(
        expand(node.left), expand(node.right))
  elif node.type=='product':
    return poly_prod(
        expand(node.left), expand(node.right))
  elif node.type=='variable':
    return {{var.label}: 1}
  elif node.type=='number':
    return {{}: node.value}
\end{lstlisting}
\vspace{-10mm}
\KTadd{\subsection{Data Structure of Product of Variables}
The products of variables in polynomials are represented by a set. In calculating the product of polynomials (i.e.,  \texttt{poly\_prod()} function in Codeblock \ref{lst:expand}), we need to calculate the product of two products of variables. This operation can be implemented by the union operation of the two input sets representing the product of variables.
In the following example, we calculate the product of $xy$ and $yz$.
\begin{align}
    xy \times yz = xyz,~ x,y,z\in \{0,1\}
\end{align}
We can confirm that the set $\{x,y,z\}$ is a union of the two sets: $\{x,y\}$ and $\{y,z\}$.
Since a polynomial is implemented as a hash map, the product of variables need to work as a key of the hash map. This means that the set must implement the  equal function and the hash function. We summarize the required implementations for a set representing products of variables.
\begin{itemize}
  \item Constructing a new set object from an object.\\ Ex) $\texttt{setA} \leftarrow \texttt{new Set(a)}$.
  \item Union operation to calculate the product.
  \item Equality operation to work as a key of hash map.
\end{itemize}
Using a debugging tool, we observed that the equal function is called the most frequently among the operations above.
}
\KTadd{
Let us consider the appropriate implementation of the set. First, we considered using the set class provided by the C++ standard library, i.e., tree-based set (\texttt{std::set}) and the hash-based set (\texttt{std::unordered\_set}). The time complexity of equality operation is $O(k)$ for the tree-based set \cite{cpp:set} and $O(k^2)$ for the hash-based set \cite{cpp:unorderedset} in the worst case, where $k$ is the size of the set. Therefore, the tree-based set is suitable in the case where the equality operation needs to run fast.}

\begin{figure*}[!htp]
\centering
\includegraphics[width=16cm]{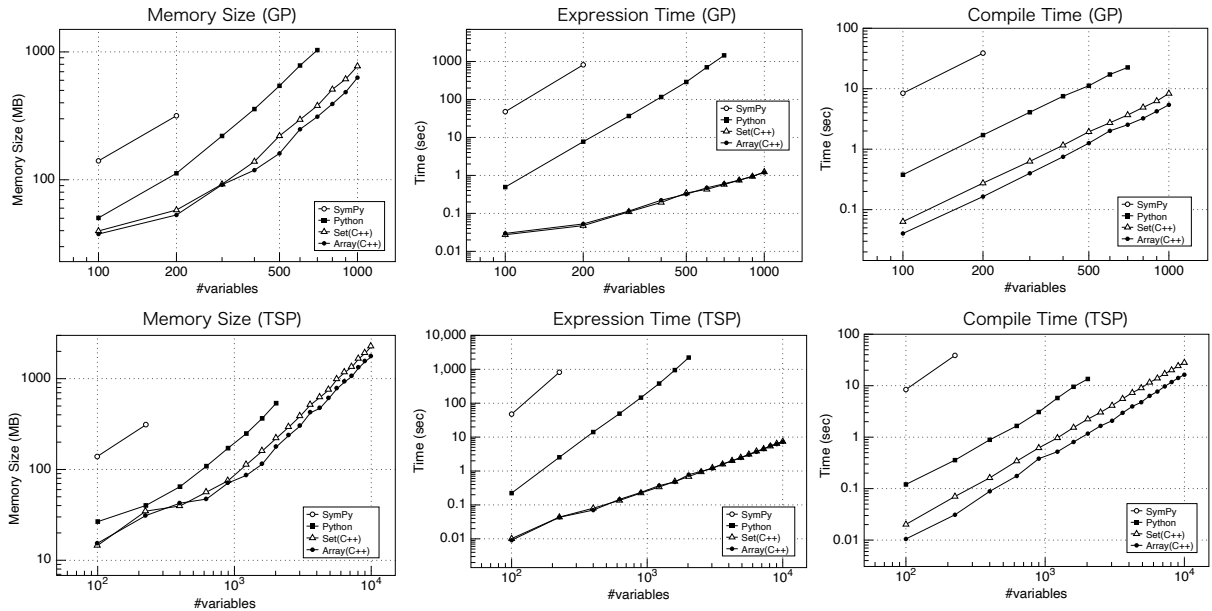}
\caption{\KTadd{Dependence of the memory size and the running time on the number of variables in QUBOs for graph partition problems (GP) and traveling salesman problems (TSP). The expression time is the running time to construct the expression, and the compile time is the running time to create the model object from the expression and produce a QUBO matrix by using the \texttt{to\_qubo()} method.}}
\label{fig:compile_runtime}
\end{figure*}

\KTadd{
Second, we considered using a sorted array to represent a set. By comparing elements from the head of the sorted array, we found that the time complexity of the equality operation is $O(k)$ in the worst case. By merging elements from the head of the array, we discovered that the time complexity of the union operation is $O(k)$ in the worst case\cite{cpp:merge}. Since the time complexity of the union operation of the tree-based set is $O(k \log k)$, the union operation of the sorted array is expected to be faster than that of the tree-based set.}
\KTadd{\subsection{Benchmark of Memory Size and Running Time}
We measured the memory size and the running time required to construct the expression and the QUBO matrix with a couple of types of combinatorial problems: graph partition problems (GP) and traveling salesman problems (TSP)\cite{lucas2014ising}. Graphs used in GP are generated as a binomial graph with the edge density set to $0.3$.
While the QUBO matrix of GP is dense, i.e., all elements of the matrix are non-zero, the QUBO matrix of TSP contains about $n_{c}^3$ non-zero elements out of $n_{c}^4$, where $n_c$ is the number of cities in TSP. We chose these two problems for the benchmark since each QUBO matrix has different characteristics in terms of the density.
In the measurement of the memory size, we measured the maximum memory size for all processes (i.e., from constructing the expression through producing the QUBO matrix). We measured the running time to construct the expression and produce the QUBO matrix separately. We compared the following implementations in this benchmarking.
\begin{itemize}
    \item \textbf{SymPy:} Using SymPy \cite{sympy} package to create QUBOs from the expression.
    \item \textbf{Python:} Older PyQUBO (version 0.4.0) implemented entirely in Python.
    \item \textbf{Set(C++):} Using tree-based set (\texttt{std::set}) for products of variables, implemented in C++.
    \item \textbf{Array(C++):} Using sorted array for products of variables, implemented in C++. Equivalent to PyQUBO (version 1.0.7).
\end{itemize}
SymPy\cite{sympy} is a well-used general symbolic tool. We used the \texttt{expand()} method of symbol objects in SymPy to create a QUBO matrix.
While PyQUBO (version 0.4.0) is implemented entirely in Python, PyQUBO (version 1.0.7) is internally implemented in C++11.
We compared two implementations of the products of variables in C++, the tree-based set \texttt{std::set} and the sorted array.
We ran our experiments on Mac OSX 10.15 and Intel Core i7 1.7GHz with 16GB memory. We used SymPy version 1.1.1 for our calculation.
We set the time limit as $3500$ seconds, and stopped calculations whose running time exceeded this duration.}
\vspace{-2mm}
\KTadd{\subsection{Result of Benchmark}
We show the dependence of memory size and running time on the number of variables in a QUBO (Fig. \ref{fig:compile_runtime}). We could not run SymPy and Python with larger problem sizes because of the time limit. We observed that the memory size and the running time of PyQUBO in C++ are better than that of PyQUBO in Python or SymPy.
We also confirmed that the memory size and the compile time of Array(C++) are superior compared to Set(C++).
In graph partition problems and traveling salesman problems, the number of terms $n$ is $m^2$ and $m^{3/2}$, respectively, where $m$ is the number of variables of QUBO. Based on the benchmark results, we can estimate the time complexity with respect to the number of terms $n$ in the Hamiltonian as shown in Table \ref{tab:order}. While the time complexity of constructing expressions in SymPy and Python is $O(n^2)$, it is reduced to $O(n)$ in C++.
Since the Python implementation requires that the sum of expressions is stored as a list, it creates a copy of the entire list when we add a new term to the expression, which leads to the complexity of $O(n^2)$.
The time complexity of constructing an expression in C++ is $O(n)$ because the expression is represented by binary trees like Fig. \ref{fig:expression_bin_tree} (left) , i.e., an added term is just referenced from the original expression, which does not require objects to be copied when we add a new term.}
\begin{table}[t]
\caption{\label{tab:order} \KTadd{The estimated time complexity with respect to the number of terms $n$ in the Hamiltonian, based on the benchmark result (Fig. \ref{fig:compile_runtime}).}}
\centering
\begin{tabular}{lcc}
\specialrule{.1em}{0em}{0.1em}
           & Expression Time & Compile Time \\ \specialrule{.1em}{0em}{0.1em}
SymPy      & $O(n^2)$   & $O(n)$ \\ [.1em]
Python     & $O(n^2)$   & $O(n)$ \\ [.1em]
Set(C++)   & $O(n)$     & $O(n)$ \\ [.1em]
Array(C++) & $O(n)$     & $O(n)$ \\ \specialrule{.1em}{0em}{0em}
\end{tabular}
\end{table}

% The compilation of expressions can be implemented with other general symbolic tools, such as SymPy\cite{sympy}. However, the compilation runtime of such program can be very long when the number of variables is large. Since PyQUBO’s compilation is optimized for QUBO creation, it runs faster than general symbolic tools. 
% The implementation of PyQUBO's compilation is optimized using the common characteristics of Hamiltonians, that is, the degree of polynomials is small (at most two in most cases), and most of the operations is sum of expressions. We compare the compilation time of PyQUBO and SymPy for the Hamiltonian of the Quadratic Assignment Problem (QAP).
%We show the dependency of the runtime on the problem size $n$, i.e. number of items in QAP in Fig.~\ref{fig:compile_runtime}. See \cite{nishimura2019item} for more details of Hamiltonians of QAP. The difference between the runtime of SymPy and that of PyQUBO becomes more pronounced as $n$ increases. We confirmed that PyQUBO’s compilation is almost 1000 times faster than that of SymPy when $n=12$. Our experiments was run on MacOS 10.13 and Intel Core i7 3.30GHz with 16GB memory. The version of SymPy used in this calculation is 1.3.

\vspace{-2mm}
\section{Conclusion}
The increasing availability of Ising machines including quantum annealing machines has been an advantage for research into practical solutions to large combinatorial optimization problems. However, Ising machines are limited to solving QUBOs and Ising model Hamiltonians, which are difficult to create and implement for many, if not most, optimization problems. The PyQUBO package offers a means of writing QUBOs and Ising model Hamiltonians in an intuitive and readable manner. Not only does it accommodate a wide variety of problems, but it also supports the automatic validation of constraints and parameter tuning, which are essential for debugging large problems. Furthermore, as an open-source library, it enables users to create and contribute new functions that are suited to their needs. We expect that many researchers will use PyQUBO as they integrate Ising machines including quantum annealing into their fields.

\ifCLASSOPTIONcompsoc
  \section*{Acknowledgments}
\else
  \section*{Acknowledgment}
\fi

One of the authors~(S.~T.) was partially supported by JST, PRESTO Grant Number JPMJPR1665, Japan and JSPS KAKENHI Grant Number 19H01553.

\ifCLASSOPTIONcaptionsoff
  \newpage
\fi

\bibliographystyle{IEEEtran}
\bibliography{main}

\begin{IEEEbiography}[{\includegraphics[width=1.0in,height=1.25in,clip,keepaspectratio]{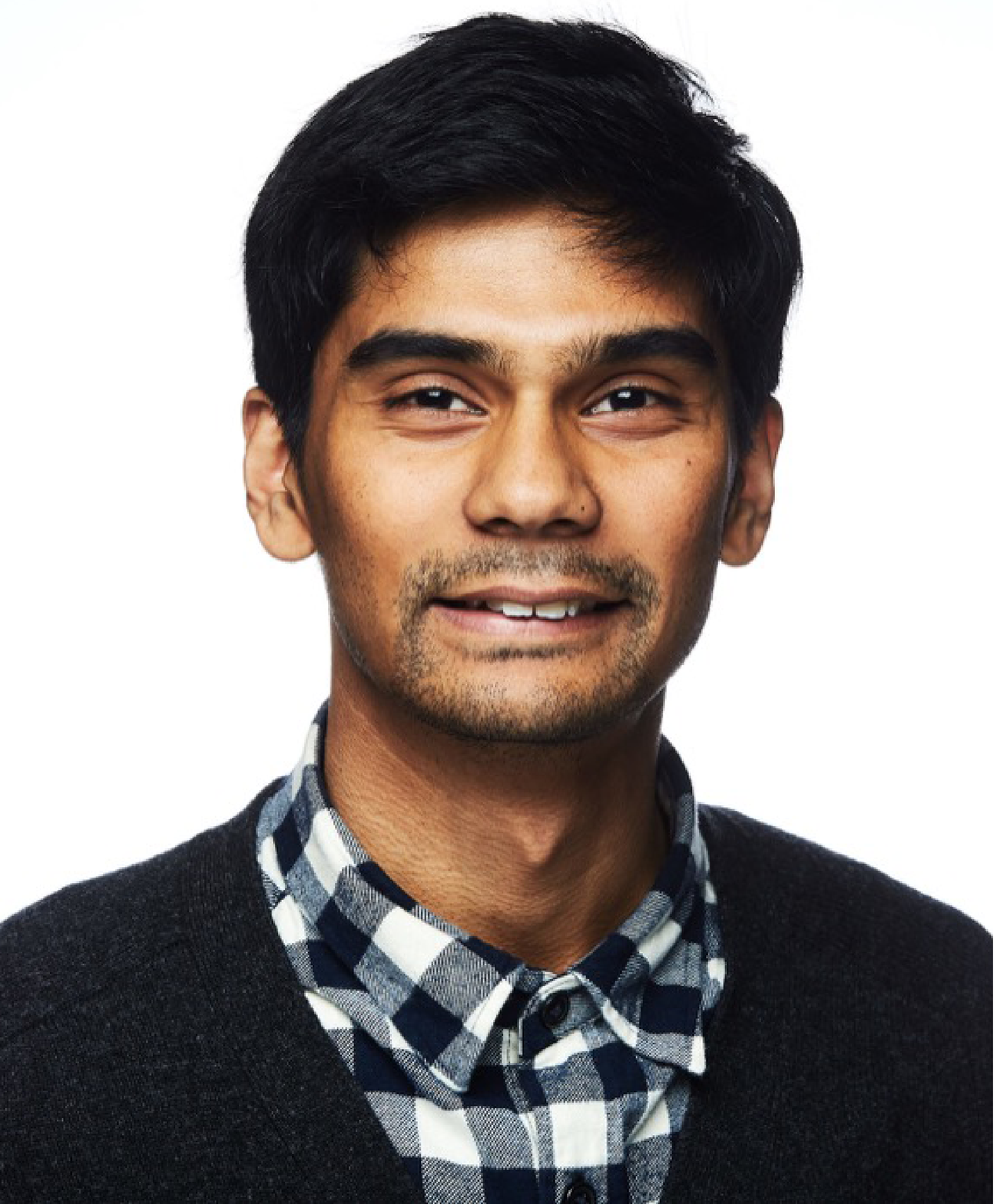}}]{Mashiyat Zaman}
received a B.A. from Amherst College in 2018. He is currently a data engineer at Recruit Communications Co., Ltd. 
\end{IEEEbiography}
\vspace{-1.0cm}
\begin{IEEEbiography}[{\includegraphics[width=1.0in,clip,keepaspectratio]{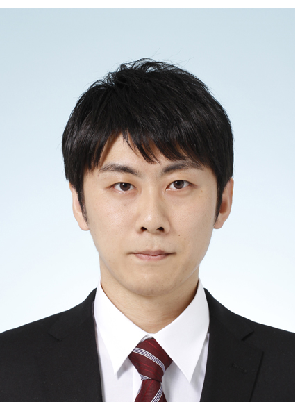}}]{Kotaro Tanahashi}
received the M.Eng. from Kyoto University in 2015. He currently works for Recruit Communications Co., Ltd. as a machine learning engineer. He is also a project manager of MITOU Target Program at Information-technology Promotion Agency (IPA). 
\end{IEEEbiography}
\vspace{-1.0cm}
\begin{IEEEbiography}[{\includegraphics[width=1in,height=1.25in,clip,keepaspectratio]{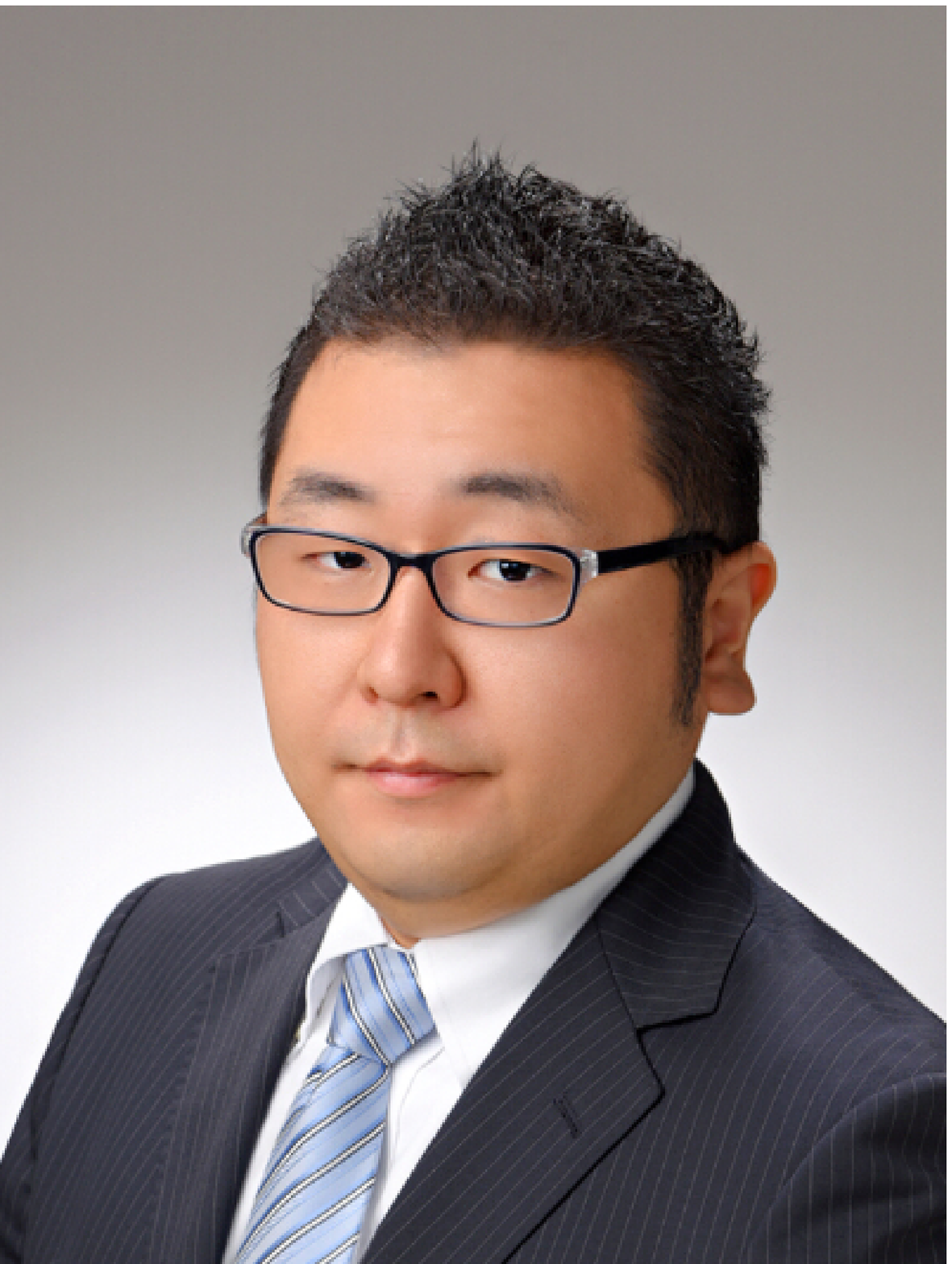}}]{Shu Tanaka}
received the Dr. Sci. degrees from The University of Tokyo in 2008. He is presently an associate professor in Department of Applied Physics and Physico-Informatics, Keio University and a visiting associate professor in Green Computing Systems Research Organization, Waseda University. His research interests are quantum annealing, Ising machine, statistical mechanics, and materials science. He is a member of JPS.
\end{IEEEbiography}

\end{document}